\documentclass[graybox, nosecnum]{svmult}

\usepackage{mathptmx}       
\usepackage{helvet}         
\usepackage{courier}        
\usepackage{type1cm}        
\usepackage{makeidx}         
\usepackage{graphicx}        
\usepackage{multicol}        
\usepackage[bottom]{footmisc}
\usepackage{hyperref}        
\usepackage{soul}            
\hypersetup{colorlinks=true,urlcolor=blue}
\usepackage[square,numbers]{natbib}

\makeindex             

\begin{document}
\title*{The Spectrometer Telescope for Imaging X-rays (STIX) on Solar Orbiter}
\author{Laura A. Hayes\thanks{corresponding author}, Sophie Musset, Daniel M{\"u}ller and Säm Krucker}
\institute{Laura A. Hayes, Sophie Musset, and Daniel M{\"u}ller \at European Space Agency (ESA), European Space Research and Technology Centre (ESTEC), Keplerlaan 1, 2201 AZ Noordwijk, The Netherlands, \email{laura.hayes@esa.int}
\and Säm Krucker \at University of Applied Sciences and Arts Northwestern Switzerland, Bahnhofstrasse 6, 5210 Windisch, Switzerland \& Space Sciences Laboratory, University of California, 7 Gauss Way, 94720 Berkeley, USA}
%
%
\maketitle
\abstract{
The Spectrometer/Telescope for Imaging X-rays (STIX) is one of the 10 instruments on-board the scientific payload of ESA's Solar Orbiter mission. STIX provides hard X-ray imaging spectroscopy in the 4-150~keV energy range, observing hard X-ray bremsstrahlung emission from the Sun. These observations provide diagnostics of the hottest thermal plasmas ($>$10~MK) and information on the non-thermal energetic electrons accelerated above 10~keV during solar flares. STIX has a spectral resolution of 1~keV, and employs the use of in-direct bi-grid Fourier imaging to spatially locate hard X-ray emission. Given that STIX provides critical information about accelerated electrons at the Sun through hard X-ray diagnostics, it is a powerful contribution to the Solar Orbiter suite and has a significant role to explore the dynamics of solar inputs to the heliosphere. This chapter describes the STIX instrument, its design, objectives, first observations and outlines the new perspectives STIX provides over the mission lifetime of Solar Orbiter.
}
\section{Keywords} 
X-rays; Solar Flare X-ray; Solar Orbiter; X-ray Imaging Spectroscopy
\section{Introduction}

\begin{figure}[h!]
\center
\includegraphics[width=0.7\textwidth]{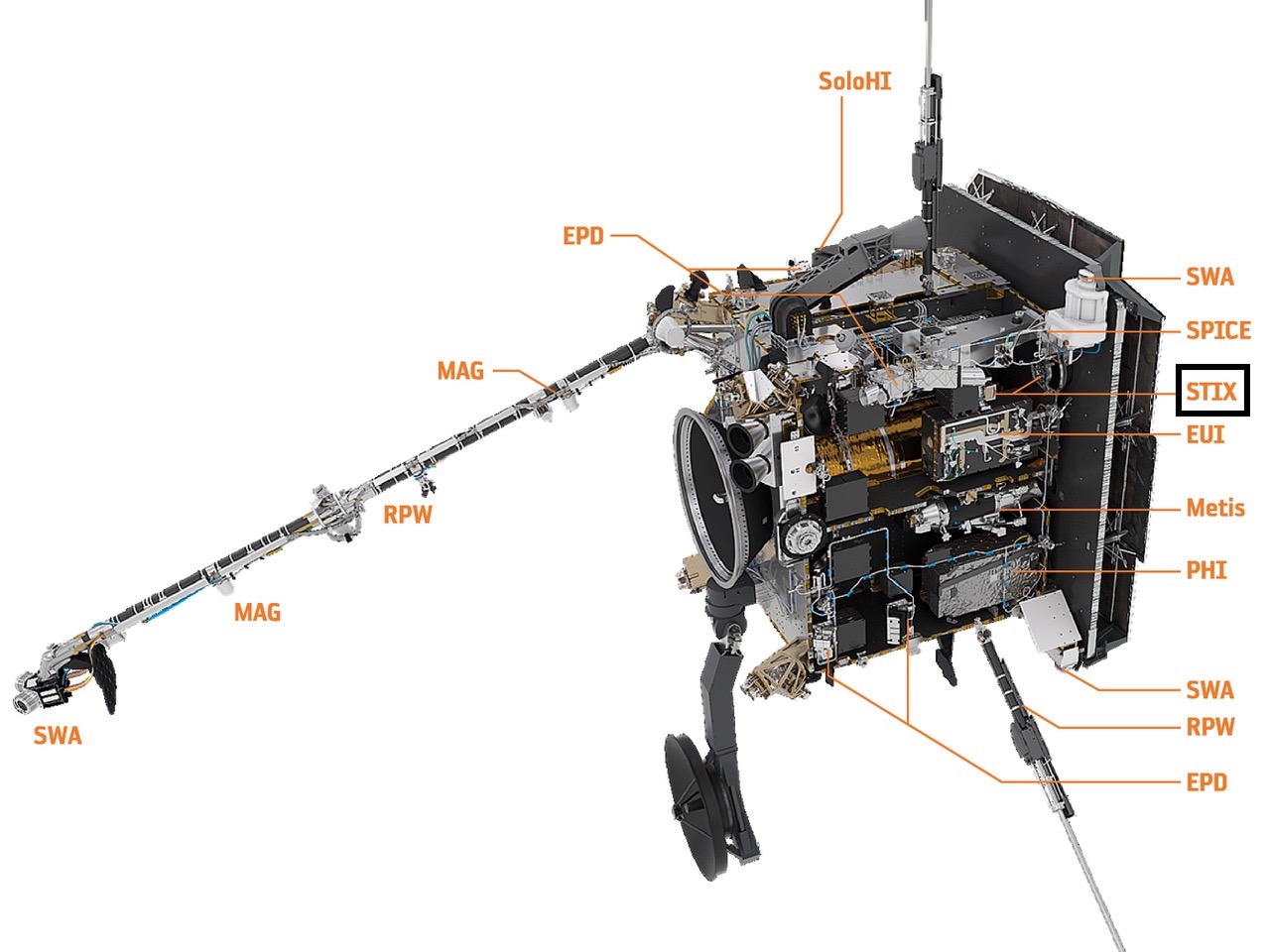}
\caption{An overview of the Solar Orbiter payload with the 10 instruments on-board. STIX is highlighted by the black rectangle. Figure from \cite{muller2020}, where more details of the individual instruments can be found.}
\label{solo_overview}
\end{figure}

\begin{figure}[h!]
\center
\includegraphics[width=0.98\textwidth]{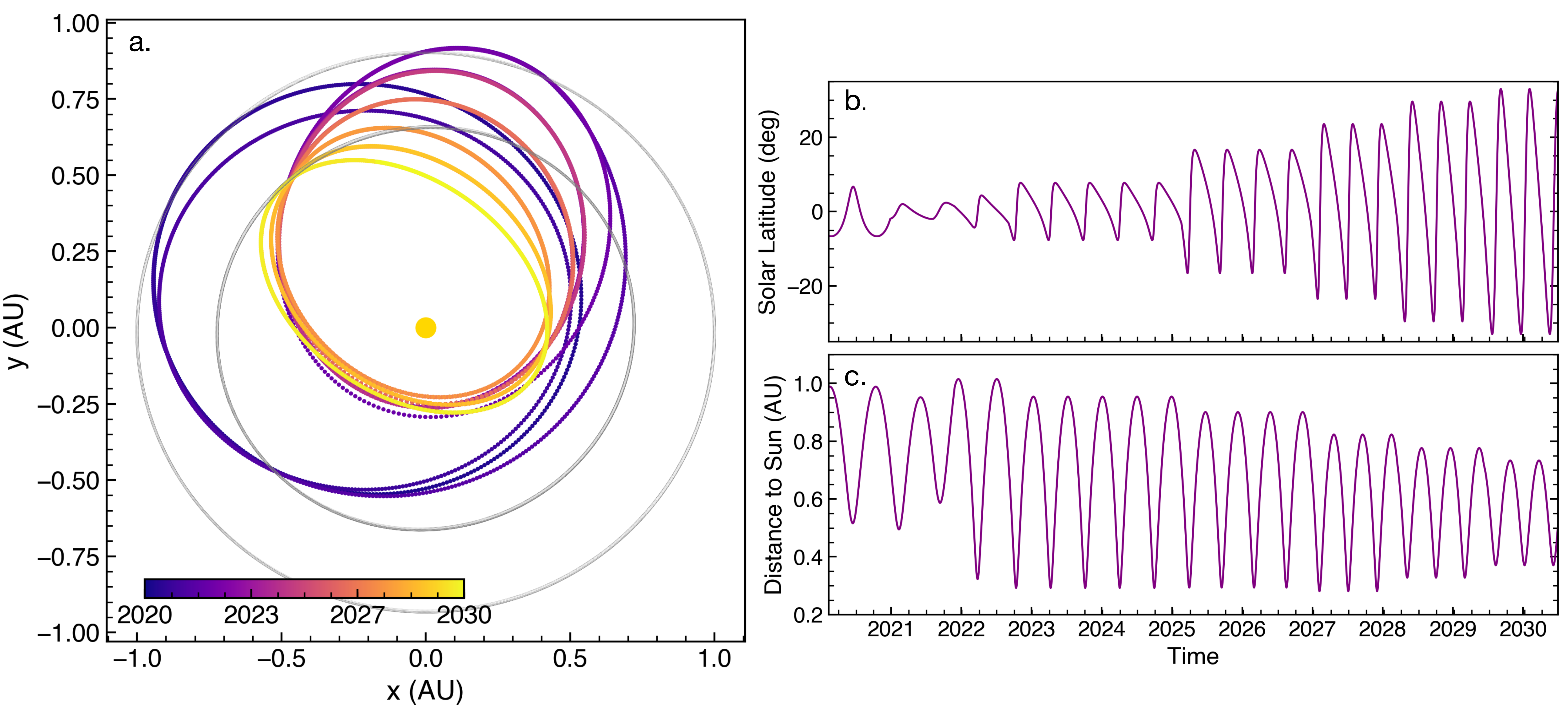}
\caption{An overview of the orbit trajectory of Solar Orbiter over the mission lifetime. The orbit as viewed from the ecliptic is shown in (a) with the colours illustrating the time since launch. The inner and outer grey circles mark the orbit of Venus and Earth, respectively. The profile of the heliocentric latitude is shown in (b) and the distance to the Sun is shown (c), both as a function of time.}
\label{solo_orbit}
\end{figure}

Solar Orbiter is a solar and heliospheric mission led by the European Space Agency (ESA) in partnership with NASA that was launched in February 2020 to investigate the interaction between the Sun and the heliosphere. It carries a comprehensive scientific payload of 6 remote sensing and 4 in-situ instruments, combining capabilities to both image and perform spectral analysis of the Sun, and measure the immediate environment of the solar wind and heliospheric conditions at the spacecraft \citep{muller2020}. An overview of the scientific payload is shown in Figure~\ref{solo_overview}. Solar Orbiter has a unique orbit; travelling close to the Sun (0.28 AU at perihelion), and out of the ecliptic plane (to over 33$^{\circ}$ heliographic latitude) allowing us to image and probe the Sun and heliospheric conditions at view-points significantly different to Earth. This orbit is achieved through a series of gravity assist manoeuvres with both Venus and Earth, with the full orbital trajectory for Solar Orbiter over the mission lifetime from launch shown in Figure~\ref{solo_orbit}. Following launch, Solar Orbiter was in a near-Earth commissioning and cruise phase until 26 November 2021 when it performed an Earth fly-by to inject into a heliocentric science orbit which marked the beginning of nominal science operations. 
 
The Spectrometer/Telescope for Imaging X-rays (STIX) is one of the remote sensing instruments on-board Solar Orbiter, and provides hard X-ray imaging spectroscopy measurements with a full Sun field of view \citep{krucker2020}. STIX measures hard X-ray emission from the Sun in the energy range of 4-150~keV, granting diagnostics of both the hot flare plasma in the solar corona and the non-thermal electrons accelerated at the Sun during solar flares and explosive energetic events. Unlike other remote sensing imaging instruments, STIX does not use mirrored focusing optics, but instead exploits an in-direct bi-grid Fourier-transform imaging technique similar to its solar predecessors; the Hard X-ray Telescope (HXT) on-board the Yohkoh mission \citep{yohkoh}, and the Ramaty High Energy Solar Spectroscopic Imager (RHESSI) \citep{rhessi}. STIX imaging is achieved by means of placing 30 pairs of X-ray opaque Tungsten grids at slightly different pitches in front of 30 coarsely pixelated Cadmium Telluride (CdTe) X-ray detectors. Incoming hard X-rays create a large scale moiré pattern on the detectors, encoding the direction and spatial extent of the incoming hard X-ray flux. The measurements from each detector are individual spatial Fourier components (visibilities) of the incident X-rays which can be used through reconstruction algorithms to create an X-ray image for a given time and energy. The grid pitches and orientations were designed to provide spatial information on angular scales on the Sun of 7 to 180''. This imaging capability, together with a spectral energy resolution of 1~keV (at 6~keV) and a dynamically-adjusted temporal resolution as low as 0.1~s allows STIX to quantify the location, intensity, spectrum, and timing of hot thermal plasma and accelerated electrons at the Sun.

The capability of STIX to measure the spatial, spectral and temporal components of hard X-rays at the Sun plays an important role on the scientific platform of Solar Orbiter. Through X-ray diagnostics, STIX provides an observational connection of the energetic electrons accelerated at the Sun to those measured in-situ at the spacecraft. In this way, STIX serves to link both remote sensing and in-situ measurements of Solar Orbiter, and aids to answer the science objectives of the mission; to understand how the acceleration of electrons at the Sun and their transport into interplanetary space and to determine the magnetic connection of Solar Orbiter back to the Sun.

In the remainder of this chapter, we describe the details of the STIX science objectives, the instrument design, the imaging concept, and present an overview of the early results and new perspectives STIX will provide over the lifetime of the Solar Orbiter mission.

\section{Scientific Objectives}

The main scientific objective of STIX is to study the acceleration of energetic electrons during explosive solar events such as solar flares, and to diagnose the spectral, spatial, and temporal properties of those accelerated electrons at the Sun. Hard X-rays are the most direct available diagnostic of solar flare electron acceleration as they are emitted by energetic electrons through bremsstrahlung emission from Coulomb collisions with ambient ions. During solar flare energy release, electrons are efficiently accelerated along newly formed closed magnetic field lines and produce X-ray emission as they interact with the ambient plasma. Non-thermal X-ray emission from electrons that are accelerated to energies $>$10~keV follow a power-law distribution and are emitted from the chromospheric footpoints in the dense layers of the lower solar atmosphere, and from the corona thought to be near the acceleration region itself. Thermal X-rays from a Maxwellian distribution of energetic electrons are emitted from the coronal magnetic loops from hot plasmas heated to temperatures $>$10~MK. Given that observed X-ray spectra from a solar flare reflect the underlying properties of the emitting electron distributions, their measurements can be used together with parametric models of bremsstrahlung emission mechanisms to infer the detailed properties of the accelerated electrons and provide quantitative diagnostics of the thermal plasma.  

While it is well known that solar flares are exceptionally efficient particle accelerators, the exact mechanism for this acceleration and the location of the acceleration region remains unknown. Moreover, the transport processes following the acceleration of the electrons both within the closed magnetic loops, and those accelerated away into the heliosphere along open magnetic field lines are not well constrained. The remote sensing X-ray measurements made by STIX provides new perspectives to further understand the electron acceleration near the Sun. Through X-ray imaging spectroscopy of well-observed flares, STIX can enable the measurements of both the temperature and emission measure of the hot thermal plasma heated by the flare, and diagnose the energy spectrum of the non-thermal accelerated electrons both from a spatially integrated and a spatially resolved perspective. The viewpoint of Solar Orbiter, which can be significantly different to Earth, means that these observations can be combined with near Earth X-ray observatories to allow us to study hard X-ray emission stereoscopically for the first time. Furthermore, by combining X-ray observations from STIX with the in-situ measurements of accelerated electrons in the heliosphere from instruments on-board Solar Orbiter, new insights about the connection and transport processes of electron acceleration both at the Sun and in interplanetary space are expected.

\section{Instrument Design and Description}

An X-ray imager designed to be part of the scientific payload of Solar Orbiter comes with limiting constrains on the size, mass, power resources, and telemetry. The use of focusing optics for X-ray imaging was hence precluded by these limitation as such instrumentation requires a large boom. An in-direct (non-focusing) grid-based Fourier imaging approach is well-suited for a X-ray payload of Solar Orbiter as X-ray imaging spectroscopy can be achieved with limited size and mass of an instrument, and has intrinsically low telemetry as the imaging information is encoded in the Fourier components of the detected counts (visibilities). STIX was developed as the X-ray scientific payload of Solar Orbiter, utilising in-direct imaging that built upon the heritage of previous solar Fourier-based imagers \citep[e.g.][]{rhessi, yohkoh}.

\begin{figure}[h!]
\center
\includegraphics[width=0.85\textwidth]{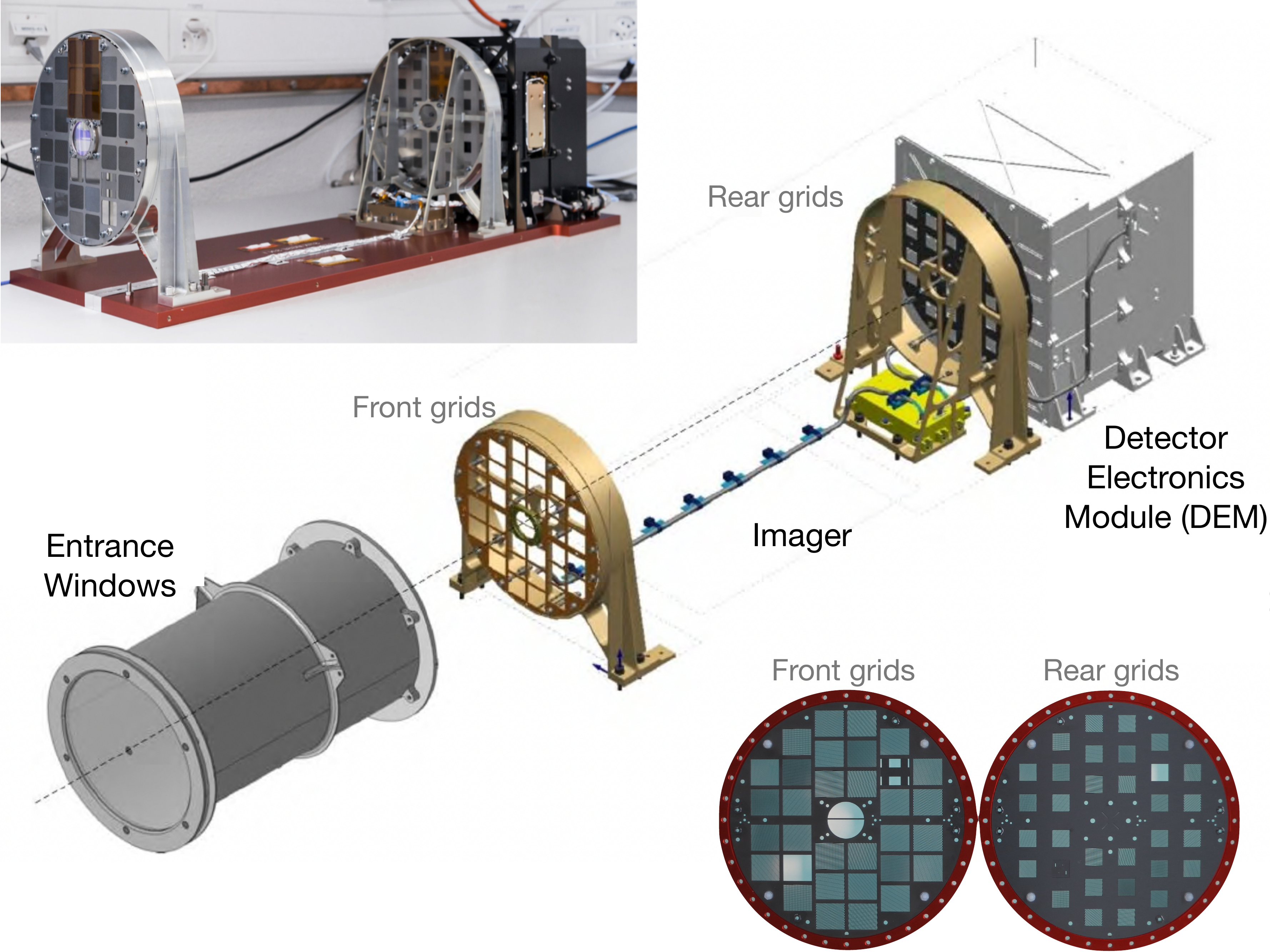}
\caption{An overview of the STIX instrument. A photograph of the STIX instrument showing the imager and the DEM is shown in the upper left hand corner, and a schematic drawing of instrument is shown in the main figure. Adapted from \cite{krucker2020}. The bottom right hand figure shows an image of the prototypes of the front and rear grids, illustrating the different grid patterns of each subcollimator, from \cite{grid_images}.}
\label{stix_overview}
\end{figure}

The STIX instrument comprises of three main parts, which are schematically shown in Figure~\ref{stix_overview}. These include an X-ray entrance window that provides thermal protection and reduces the intense flux from entering the instrument when close to the Sun, the imaging system that contains the grid imaging hardware and the aspect system, and the detector/electronics module (DEM) which contains the X-ray CdTe detectors, a movable X-ray attenuator, and the electronics.  Over the next few sections, the details of these components are discussed in detail.

\subsection{The Entrance Window}
Solar Orbiter will fly as close as 0.28~AU to the Sun, and hence high solar flux must be prevented from entering the spacecraft. The entrance window to the STIX instrument consists of two X-ray transparent Beryllium windows that are independently mounted on the Solar Orbiter heat shield. The front and rear window have a thickness of 2~mm and 1~mm, respectively, with the front window coated with the same {\it SolarBlack}\footnote{\tt http://www.enbio.eu/solar-orbiter/} as the Solar Orbiter heat shield. The requirement for the windows are two-fold; firstly they serve to limit the incident optical and infrared solar flux into the spacecraft, and secondly they block the lower energy X-rays ($<$4~keV) that would otherwise cause excessive count rates during flares and contribute to pulse-pile up and detector dead-time issues. There are two small openings in both the front and rear entrance windows that are used for the aspect system, which is described in more detail below.

\subsection{Imaging Concept}
\begin{figure}[h!]
\center
\includegraphics[width=0.8\textwidth]{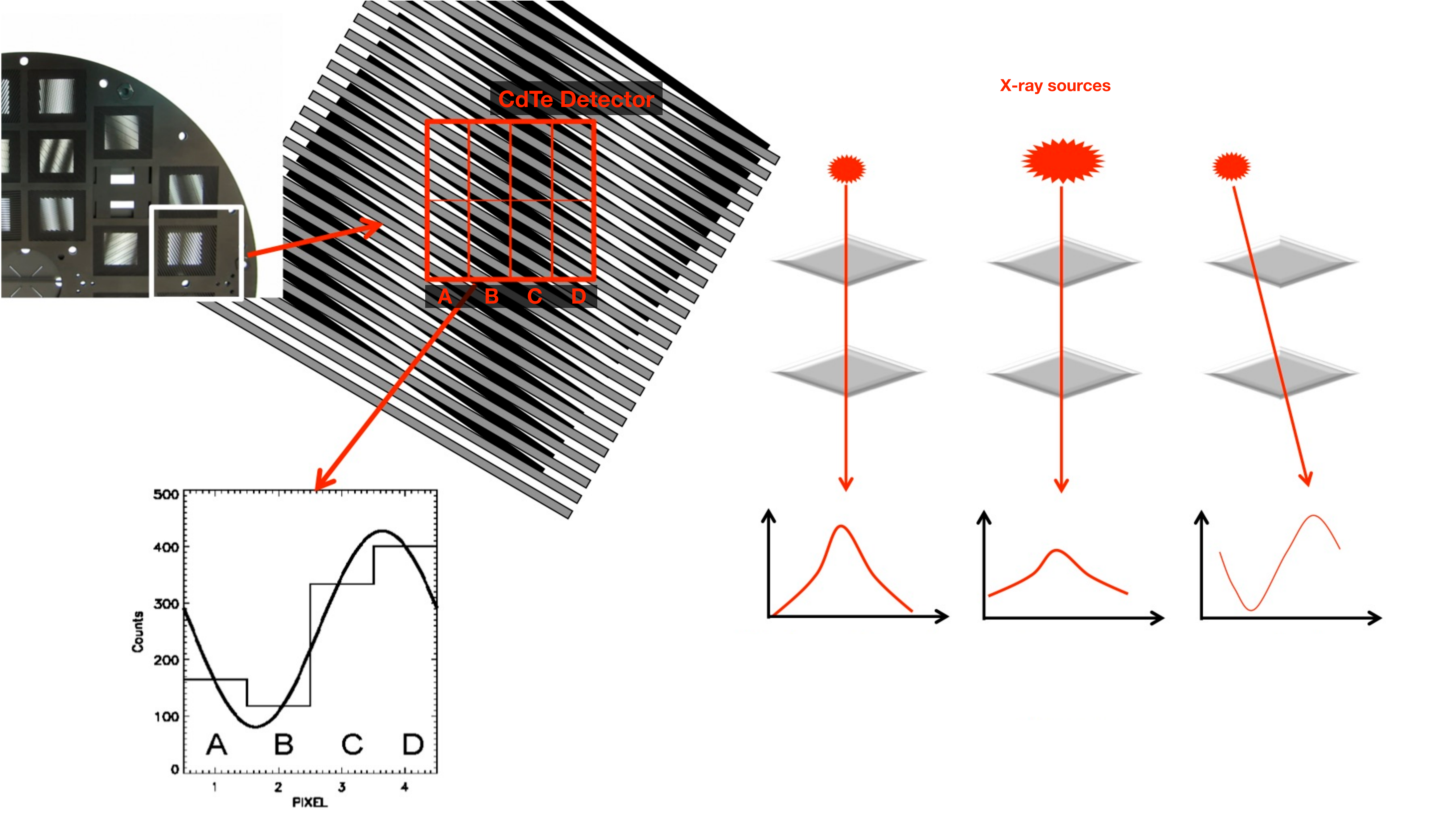}
\caption{A schematic to demonstrate a moiré pattern from a single subcollimator, and how the coarsely pixelated CdTe detectors measure the visibility. Figure modified from \cite{moire_pattern_image}.}
\label{moire_pattern}
\end{figure}

To briefly explain the concept of in-direct bi-grid Fourier imaging, we can consider an X-ray source that is viewed simultaneously through multiple `subcollimators', each of which measures one spatial Fourier component of the X-ray source distribution. The term ‘subcollimator’ refers to a pair of widely separated X-ray opaque grids in front of an X-ray photon detector. Incident X-rays on the subcollimator will be selectively absorbed or transmitted by the pair of grids depending on the direction of the source to form a modulation pattern on the detector \citep[see][]{prince}. In the case of STIX, the front and rear grids contain a large amount of parallel, equally spaced slits, and each subcollimator is designed such that there is a small difference in the pitch and/or orientation of the grids, which produces large-scale periodic moiré patterns on each detector. The amplitude and phase of the moiré pattern provides a direct measurement of a spatial Fourier component of the angular distribution of the source, also known as a visibility. STIX uses coarsely pixelated CdTe detectors behind the grid pairs such that the moiré pattern for each detector is detected through the count rate differences between the pixels. A schematic example of this is shown in Figure~\ref{moire_pattern}. Here we can see how an individual subcollimator measures a visibility. The pair of grids form a periodic moiré pattern, which is then seen by the CdTe detector in the four pixels in a row (A, B, C, D). Differences among the count rates of the pixels allow us to calculate the amplitude and phase of the visibility. The right hand side shows a schematic of how different X-ray source distributions change the detected characteristic values of the measured visibility.

There are 32 subcollimators on STIX, 30 of which are used for imaging. The distance between the grids, and the average pitch and orientation of the pair of grids, determine the frequency and orientation of the Fourier component of the incident X-ray source distribution that is measured by the subcollimator \citep{giordano}. At a given time and energy, an image can be reconstructed by inverting the Fourier transform of the data of the photon flux, which is represented by the complex visibilities sampled in the spatial frequency domain (also known as the (u, v) plane). Such visibility measurements are analogous to the imaging information provided by antenna pairs in radio interferometry. This in-direct Fourier based technique to image X-rays with STIX has the advantage that it can combine a large full Sun field of view with a fine angular resolution as small as 7’’, such that a flare can be observed and analysed with imaging spectroscopy no matter where on the solar disk it occurred.

\subsection{Imaging System }
To achieve the imaging concept described above, the STIX imaging hardware consists of two X-ray opaque Tungsten grid assemblies separated by a distance of 55~cm (see Figure~\ref{stix_overview}). The grid assemblies are independently mounted on an internal spacecraft panel and placed in front of the CdTe detectors located in the DEM. The grids themselves are 400~$\mu$m thick, and have pitches ranging between 38~$\mu$m - 1~mm. The X-ray measurements at the detectors from the subcollimators with different pitches provide the different angular Fourier components, or visibilities, that sample the (u, v) plane. The pitches are logarithmically spaced to allow for a wide range of source angular sizes to be resolved. These pitches correspond to angular spatial resolutions of 7’’ to 180’’ on the Sun.

STIX creates 30 Fourier components of an X-ray source distribution. These components, or visibilities, can be used to reconstruct images at different photon energies by means of various computation algorithms that invert the Fourier transform from the limited data. Image reconstruction from measured visibilities has a heritage in radio interferometry, and many techniques can be leveraged in the X-ray domain for a similar purpose. These methods include simple back-projection, CLEAN, forward-fitting, count-based methods, among others \citep[e.g.][]{perracchione, massa_vis}. The advantage of in-direct imaging, is that the image data is downlinked as visibilities that have very low telemetry, and then the data processing and image reconstruction is done on the ground. One of the disadvantages however, is that given the relatively few visibilities available with STIX, there is a limitation on the complexity of the source that can be imaged. 

There are two subcollimators that are not used for imaging. One of these is the background detector that is used for the continuous monitoring of the X-ray background, and also provides low energy X-ray measurements during large flares as the X-ray attenuator does not move in front of the background detector. The background subcollimator has a fully open front grid, and a fully opaque rear window with six small apertures of varying size of 0.01, 0.1 and 1 mm$^2$ to allow certain pixels on the detector to measure flux. The remaining subcollimator is used as a coarse flare locator which allows for the real-time estimation of the flare location relative to STIX to be determined. This is achieved by having the front grid with a coarse `H'-shape pattern, and the rear grid open. This geometry allows a potential flare location to be estimated on the Sun based on the unique pixel pattern created on the detector.  

\subsubsection{Aspect System}

\begin{figure}[h!]
\center
\centerline{\hspace*{0.015\textwidth}
               \includegraphics[width=0.5\textwidth,clip=]{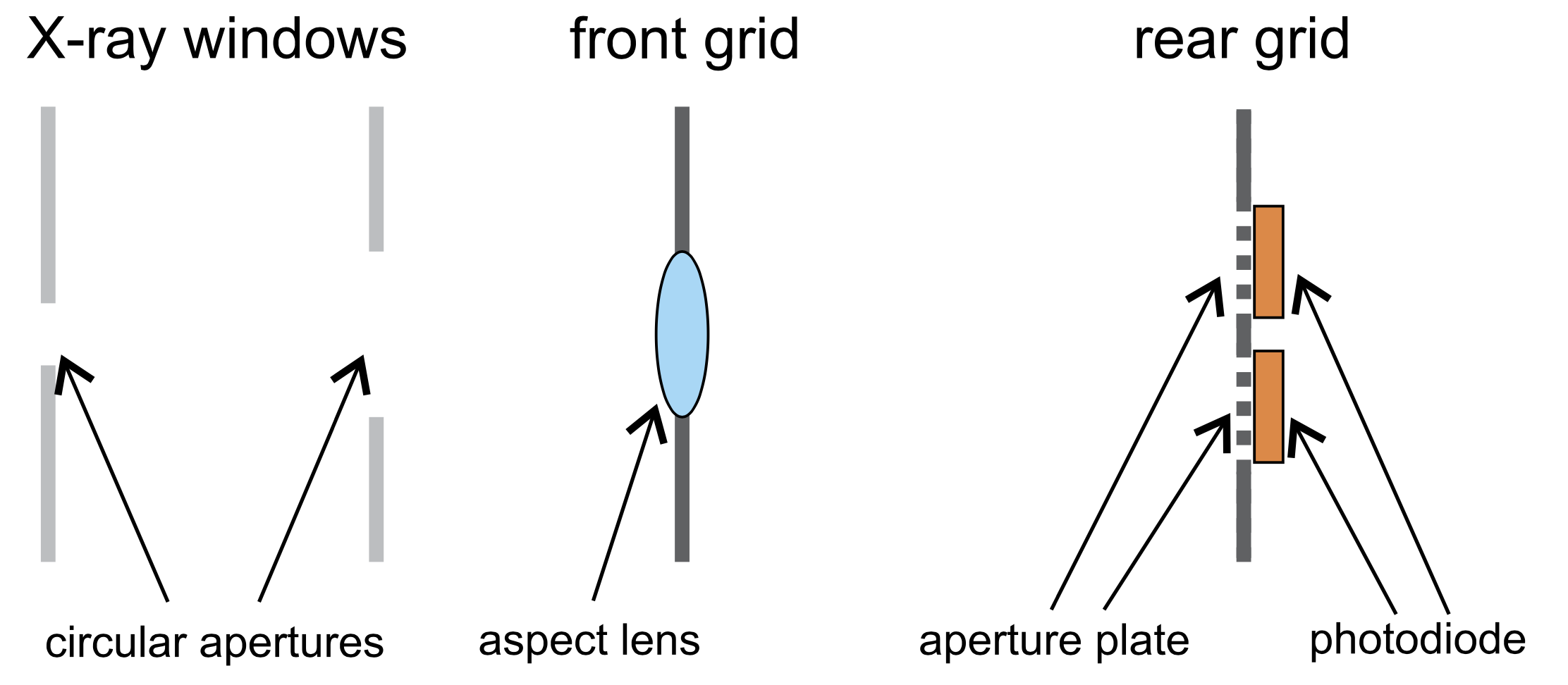}
               \hspace*{0.05\textwidth}
               \includegraphics[width=0.4\textwidth,clip=]{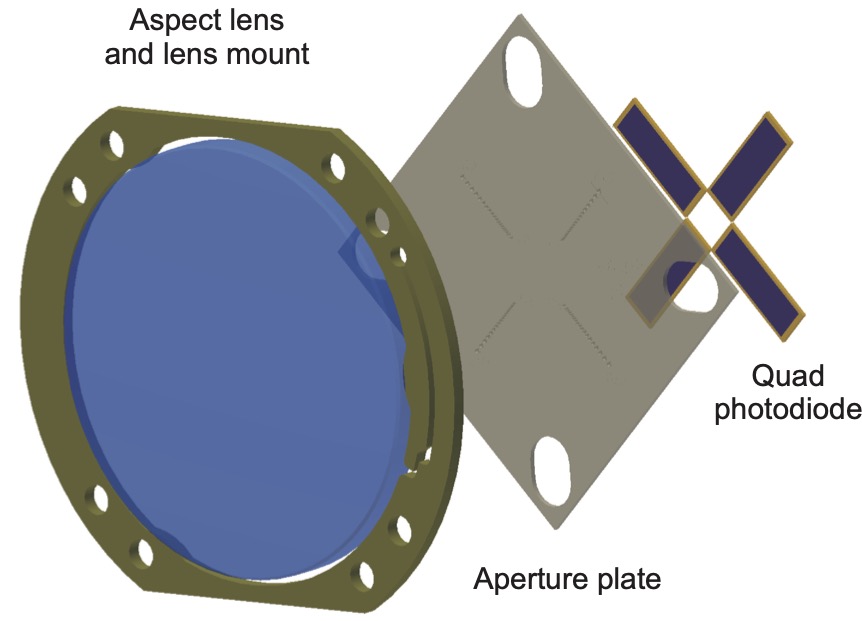}
              }
\caption{Overview of the STIX aspect system from \cite{sas}. Left: Schematic side view of the elements of the STIX aspect system. Right: CAD design renderings of the three major components: aspect lens and lens holder, aperture plate, and quad photodiode (only the four active areas are shown).}
\label{fig:sas}
\end{figure}

The pointing accuracy provided by the Solar Orbiter spacecraft itself is not sufficient to determine the absolute position of the reconstructed X-ray source on the Sun. Hence STIX requires its own aspect system to correctly determine the pointing and to allow us to place an X-ray image on the correct location of the solar disk. This is required scientifically so that STIX observations can be coordinated with other remote sensing instruments at different wavelengths, which is particularly important for solar flare studies. The STIX aspect system (see Figure~\ref{fig:sas}) is implemented as a solar limb detector and consists of a 28~mm diameter lens that is mounted in the front grid assembly that focuses an optical image of the Sun in the 550-700~nm wavelength range onto the centre of the rear grid assembly. The rear grid is perforated by a set of small apertures arranged in a cross-shaped configuration of four radial arms, and the light passing through is detected by four photodiodes behind. When the solar limb crosses one of the apertures, a step-wise change in the measured lightcurve is observed. Between the steps, a more gradual variation of the signal is caused by the limb-darkened intensity profile of the solar disk. Through measuring the time evolution of the four signals provided by the photodiodes, the direction of the STIX imaging axis with respect to the Sun centre can be determined. With this, an image placement accuracy of 4’’ can be achieved. See \cite{sas} for more details on the STIX aspect system.

\subsection{Detector/Electronics Module (DEM)}

The DEM is located behind the imaging grid system and contains all the active instrument electronics. It consists of both the Detector Box that encloses both the cold-unit containing the CdTe detectors and the movable attenuator, and the Instrument Data Processing Unit Box, which contains the digital electronics and power supply.

\subsubsection{Detectors}
\begin{figure}[h!]
\center
\includegraphics[width=0.6\textwidth]{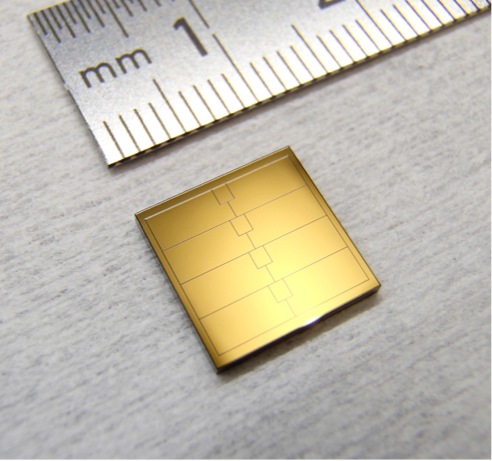}
\caption{An image of one of the STIX detectors from \cite{krucker2020}. The pixel pattern can be seen with 8 large pixels and 4 small pixels.}
\label{detector_picture}
\end{figure}

To detect X-rays and perform imaging spectroscopy in the energy range of 4-150~keV, the STIX detector system consists of an array of 32 CdTe passively cooled semiconductor detectors. The CdTe detectors are 1~mm thick and have a 10$\times$10~mm$^2$ area. The detectors are bonded to hybrid circuits called Caliste-SO, which integrate the front end read out ASIC \citep{meuris2012, gevin}\footnote{developed by CEA-Saclay}.  Each detector are coarsely pixelated into 8 large and 4 small pixels, as shown in Figure~\ref{detector_picture}. For each collimator, the difference of pitch and orientation between the front and rear grid is chosen in order to produce a moiré pattern aligned with the rows of pixels on the detector, and with a period of 8.8 mm, corresponding to the active width of the detector. A line of four pixels can therefore be used to sample the moiré pattern (see Figure~\ref{moire_pattern}). The pixel count rates as a function of energy are used to determine the amplitude and phase of the moiré pattern, and thus the Fourier components. The two lines of four large pixels in the STIX detectors provide two redundant measurements of the moiré pattern, and the smaller pixels can help address pulse pile up and the large dynamic range in expected count rates. Pulse pile-up occurs when two or more photons arrive almost at the same time such that they are detected as a single event with an energy equal to the sum of the photon energies. Smaller pixels will be less affected and hence comparisons between the larger and smaller pixels can estimate the issue during large flares.

\subsubsection{X-ray Attenuator}
As the STIX detectors are designed to observe a wide range of solar flares from the smallest micro-flares to the largest X-class flares, a mechanical attenuator is required to be inserted during times of excessive count rates to avoid detector saturation. The STIX attenuator consists of 0.6~mm thick Aluminium blades, and when inserted they preferentially reduce the lower energy X-rays to control the count rates. The attenuator is inserted and removed autonomously in front of the detector field of view within two seconds based upon the the on-board Rate Control Regime (RCR) algorithm. The design of the attenuator is such at minimum sacrifice to the higher energy photons. The attenuator is not placed in front of the background detector, and hence allows us to monitor the lower energy flux with some of the detector pixels that is blocked out from the other detectors by the attenuator during large flares. In addition to the attenuator, another rate control mechanism is to take advantage of the pixelated detectors, and utilise only the smaller pixels during times of excessive count rates during large solar flare events.

\subsubsection{On-board binning}
Given that Solar Orbiter has limited telemetry due to its orbit, the STIX data are binned on-board into 32 selectable energy bins and dynamically adjusted time bins. The energy binning is chosen at energies that are optimised for typical solar flare spectra, with a 1 keV bin width at the lower energy with gradually larger bins at higher energies. The time bins can be dynamically adjusted depending on count rate. During flares, these bins can be adjusted as low as 0.1~s. To save space in the onboard STIX memory, STIX is most frequently run with a setting down to 1~s or 0.5~s. Tests at 0.3~s has been run successfully, while faster cadences need further investigation as large flares lasting hours at high cadence might cause problems with overwriting buffers. The overall telemetry of STIX is very low, both due to this binning and due to the intrinsically low nature of in-direct imaging. For example, a STIX image in the form of visibilities can be downlinked with as low as 100 bytes.

\subsubsection{Calibration}

\begin{figure}[h!]
\center
\includegraphics[width=0.9\textwidth]{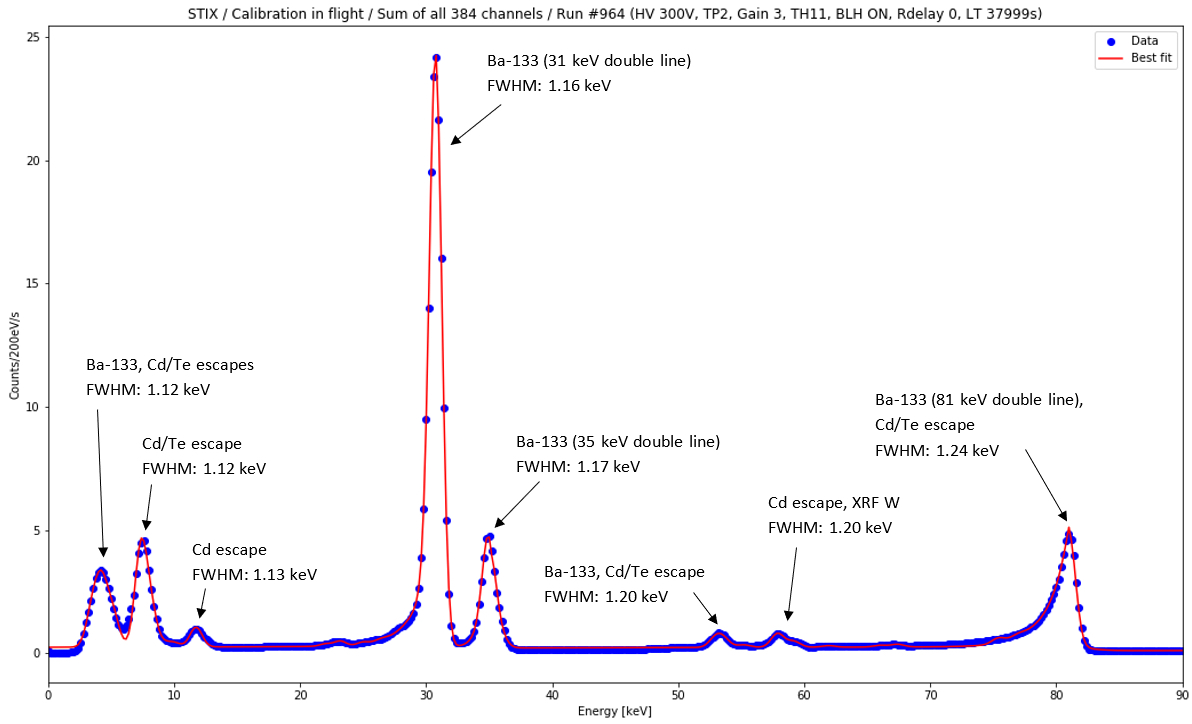}
\caption{Spectrum of the on-board STIX calibration source measured in-flight on 27 May 2020}
\label{calibration_source}
\end{figure}

The imaging technique employed by STIX requires accurate knowledge of the relative count rates between the different pixels and detectors within different energy ranges. Hence accurate calibration is required. This is achieved through use of on-board radioactive calibration sources. Each pixel is illuminated with X-rays produced from $^{133}$Ba sources, and during non-flaring times, the spectra of these sources is taken on a pixel basis and transmitted to Earth daily. The spectrum of the on-board calibration source is shown in Figure~\ref{calibration_source}, with the two strongest emission lines are 31~keV and 81~keV clearly visible, however the entire shape of the spectrum is used for calibration \citep[e.g.][]{maier}. This allows ground-based analysis to be performed to determine the updated calibration constants which are then uploaded to the STIX instrument.  The design of the source was chosen such that it does not increase the desirably low background of STIX that allows the sensitivity to weak flares, however as can be seen some CdTe escape lines occur at the lower energies which can affect some of the lower energy bins. The source has a half-life of 10.5~years, and the total source activity is 4.5~kBq.

\section{First Scientific Results \& Future Potential}

Although it has only been a short time since the beginning of the nominal science phase of Solar Orbiter which began in Nov 2021, STIX has already provided many new scientific observations and has observed thousands of solar flares ranging from small micro-flares to large X-class flares. Thanks to the low telemetry of STIX observations, many measurements were taken throughout the Solar Orbiter cruise phase, and STIX has been observing the Sun almost continuously since January 2021. Here we highlight some of the first scientific results provided by STIX and discuss the future potential further observations STIX will bring to solar flare studies over the next several years.

\subsection{First results from Cruise phase}

\subsubsection{Micro-flare Observations}
As the commissioning phase of Solar Orbiter coincided with relatively low solar activity, it provided an opportunity to study the smallest of solar flares. Some of the first results of micro-flare observations with STIX during the cruise phase are presented in \cite{battaglia2021_micro} and  \cite{saqri2022}.

\begin{figure}[h!]
\center
\includegraphics[width=0.75\textwidth]{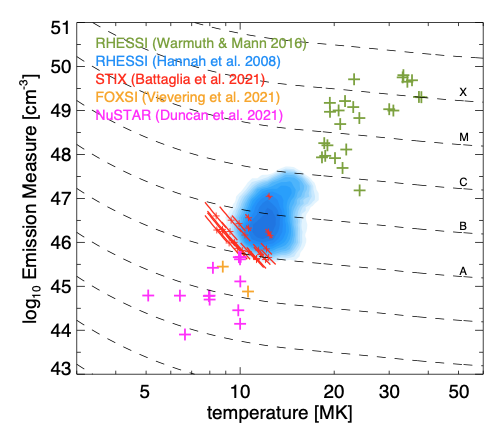}

\caption{STIX contribution to the comparison of the emission measure and temperature distributions of solar micro-flares compared with previous published observations with prior instruments. The STIX events where taken during the commissioning phase of STIX in June 2020.}
\label{microflare_dist}
\end{figure}

The authors of \cite{battaglia2021_micro} analysed a sample of microflares observed with STIX in June 2020 and used a simple isothermal flare temperature estimate  to calculate the emission measure and temperatures. The results of this are shown in Figure~\ref{microflare_dist} with the comparisons with prior observations similarly represented. As illustrated, the STIX microflare observations lie on the lower end of the prior RHESSI observations, but are not as sensitive to observations taken with focusing optics instruments such as NuSTAR and FOXSI. These observations were taken when Solar Orbiter was at 0.52~AU, so the increased sensitivity as compared to observations at Earth is a factor of two. As Solar Orbiter approaches a closer distance to the Sun during perihelion (0.28~AU), STIX's sensitivity to measure the smallest flares increases further. 

\subsubsection{First imaging results from different perspectives}
Over the Solar Orbiter cruise phase, STIX observed many flares from which the imaging capabilities of the instrument could be tested, and new studies performed with the observations. Here we highlight the first STIX imaging results published in \cite{massa2022} that illustrates the X-ray observations of a selection of solar flares as viewed from Solar Orbiter. One of the main novelties STIX provides as an X-ray imaging spectrometer is that it gives us for the first time different viewpoints of flares compared to the Sun-Earth line. This presents a new perspective, but also requires careful consideration to be taken into account when comparing images taken near Earth to those taken at the viewpoint of Solar Orbiter. As there are currently no solar-dedicated hard X-ray imager imaging the Sun at Earth, the STIX first observations can be compared with extreme ultraviolet (EUV) images from the Atmospheric Imaging Assembly on-board the Solar Dynamics Observatory (SDO) to validate the morphology of the X-ray source in relation to the flaring structure.

\begin{figure}[h!]
\center
\includegraphics[width=\textwidth]{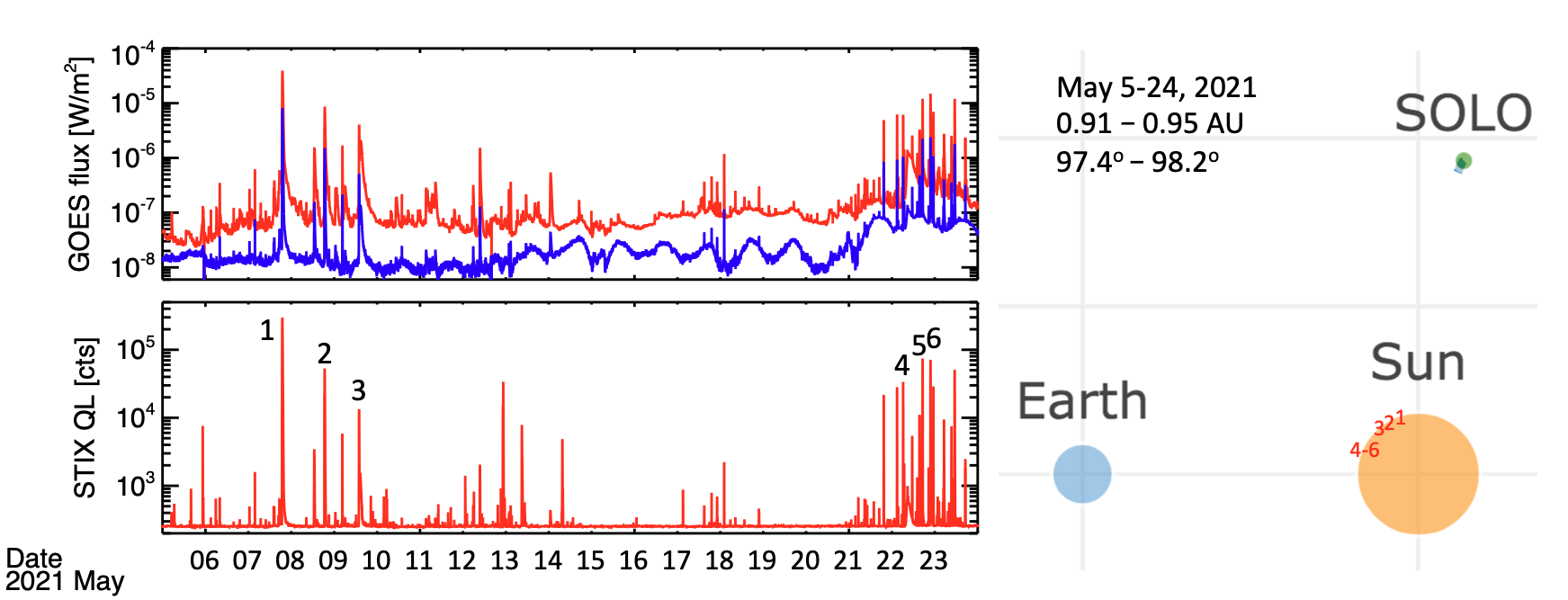}
\includegraphics[width=\textwidth]{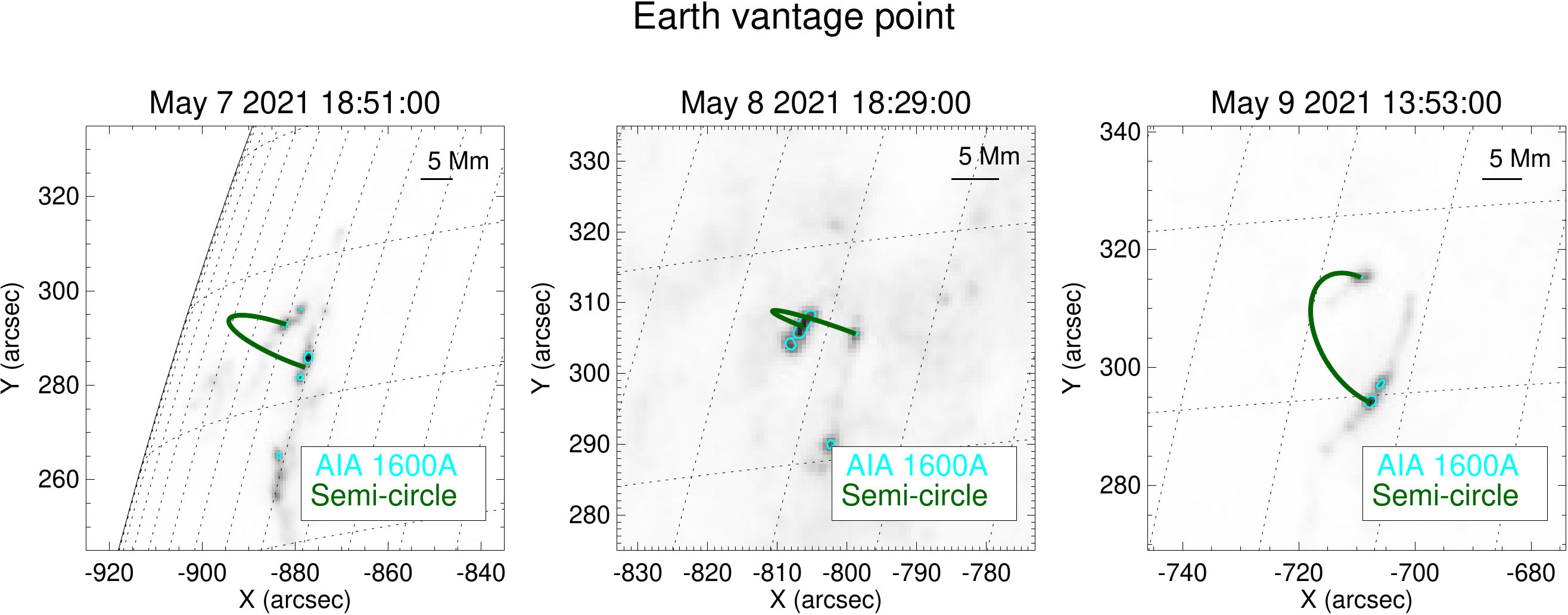}
\includegraphics[width=\textwidth]{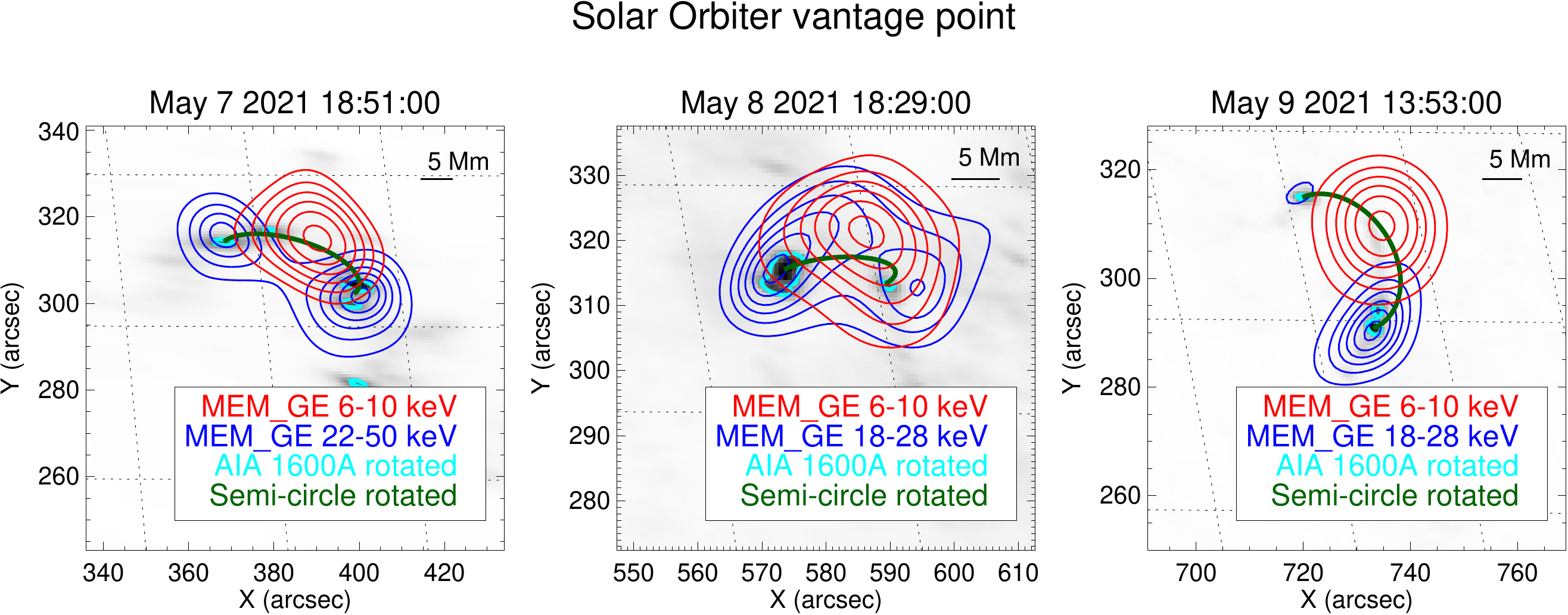}
\caption{Some of the first results of imaging different viewpoints of solar flares from both near Earth and Solar Orbiter/STIX published in \cite{massa2022}. The top panel shows the soft X-ray observations from the X-ray sensor on-board GOES, and the STIX count rate over the observing period. The flares of interest are numbered. The right hand side of the top panel illustrates the position of Solar Orbiter as relative to Earth, and demonstrates the different viewing angles of the flare events. The middle panel shows the flare as observed from the viewpoint of the Solar Dynamics Observatory (near Earth), with a plotted vertical semicircle (green) connecting the flare ribbons. The bottom panel shows the X-ray observations from STIX, where both the non-thermal footpoints (blue contours) and the thermal loops (red contours) can be clearly identified.}
\label{imaging}
\end{figure}

The results of this are shown in Figure~\ref{imaging}. The reconstructed X-ray images are overlaid on the 1600~\AA\ channel of AIA, that has been re-projected to the view of Solar Orbiter. Of the three flares demonstrates here, the Maximum Entropy Method (MEM) was used to reconstruct the X-ray images from the measured visibilities, and the non-thermal hard X-ray footpoints can clearly be distinguished from the thermal loop-top source, and the sources align well with the flaring ribbon structures seen with AIA.

\subsubsection{Imaging Spectroscopy with STIX}
\begin{figure}[h!]
\center
\includegraphics[width=0.98\textwidth]{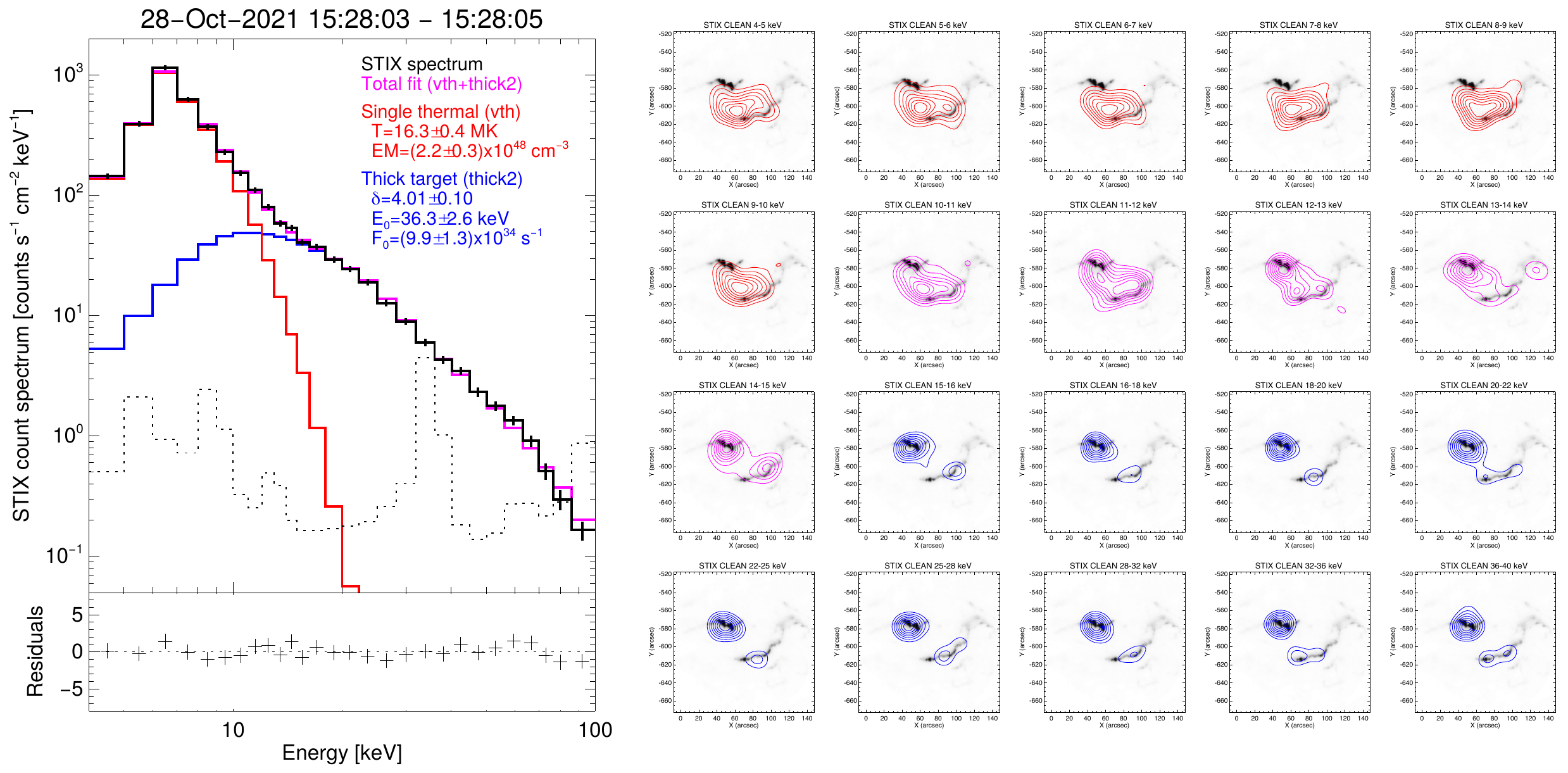}
\caption{First results demonstrating STIX's imaging spectroscopy capabilities. The left hand plot shows the STIX count spectrum that is fit with both a thermal and non-thermal emission model. The grey dashed line shows the background spectrum, which is dominated by the calibration source, and is subtracted from the spectra before analysis. The reconstructed images over the STIX energy bins are shown on the right illustrating the morphology over the energy range from the coronal hot loops at lower X-ray energy to the chromospheric footpoints at higher energy observations. Figure from \cite{rhessi_nug}.}
\label{oct_event2}
\end{figure}

Given that STIX is an imaging spectrometer, it has the capability to image at different spectral energies, and allow us to determine the spatial components of the hard X-ray spectra and distinguish the hot flare plasma sources from the non-thermal footpoints. A demonstration of STIX's capabilities is presented in \citep{rhessi_nug} for the first X-class solar flare that was observed by STIX since the launch of Solar Orbiter. In fact this was the first X-class flare of Solar Cycle 25.  In Figure~\ref{oct_event2} both the X-ray spectral analysis is demonstrated (on the left hand side) and then the X-ray imaging spectroscopy to show images reconstructed in different energy bands over the same time-interval presented in the right hand side of the figure.

\subsection{A Stereoscopic Potential - Measuring X-ray Directivity}

\begin{figure}[h!]
\center
\includegraphics[width=0.6\textwidth]{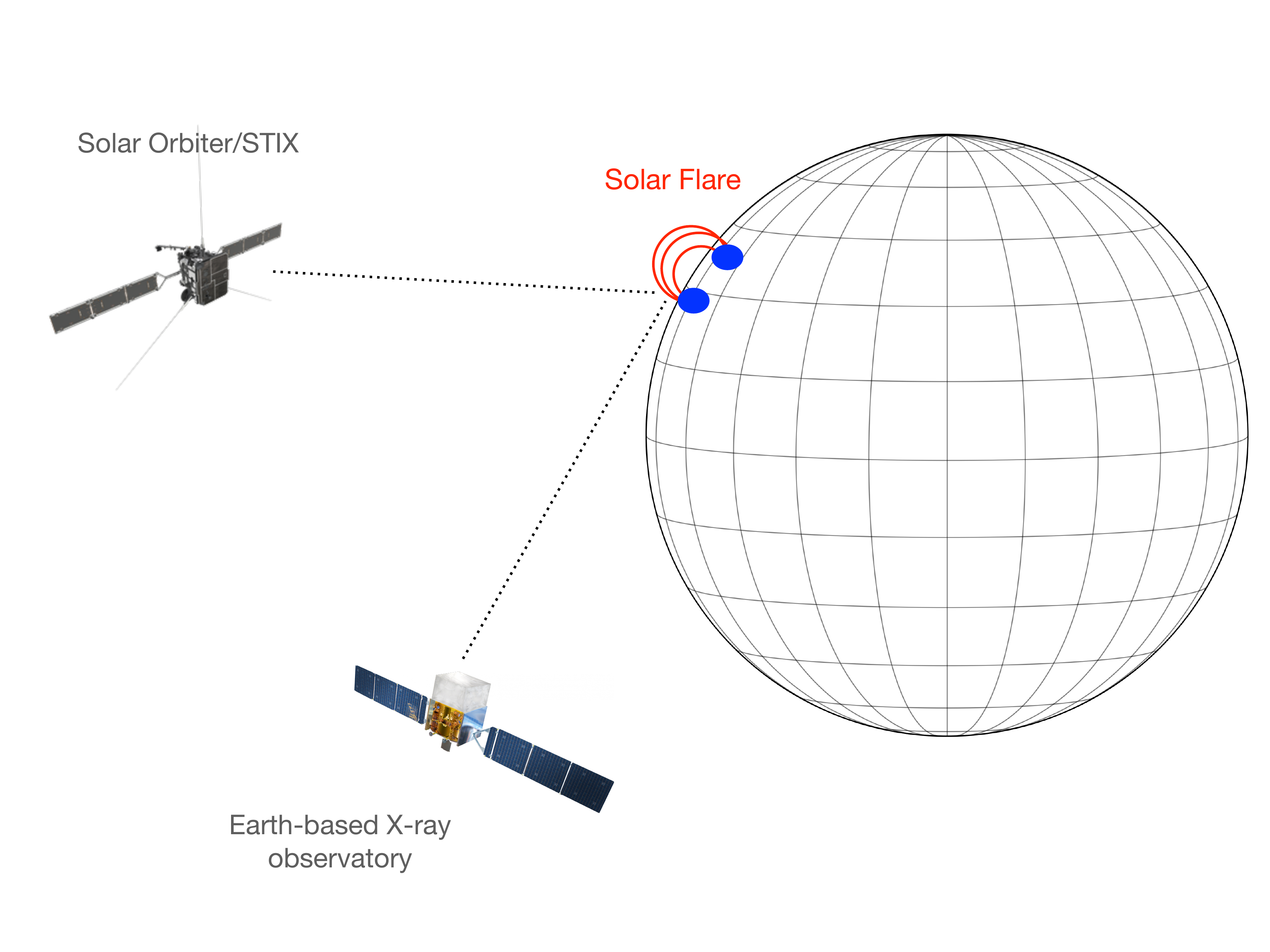}
\caption{Solar Orbiter/STIX will provide a significantly different view of solar flares than near Earth observatories providing an opportunity to observe X-ray directivity. }
\label{directivity}
\end{figure}

The trajectory of Solar Orbiter around the Sun allows us to observe the X-ray emission from flares from significantly different heliographic angles from the Sun-Earth line. By combining STIX observations with near Earth X-ray observatories, a unique opportunity is provided to perform systematic X-ray stereoscopic measurements for the first time, see cartoon Figure~\ref{directivity}. Stereoscopic hard X-ray measurements are extremely important as they can provide new information on both the angular distribution of accelerated electrons, and furthermore when imaging is available, allow us to study the coronal hard X-ray source together with hard X-ray footpoint emission which is generally not possible from one vantage point due to the extremely large instrumental dynamic range this would require.

As discussed earlier, observations of hard X-ray spectra during flares provide information about the underlying accelerated electrons distribution.  However one property that cannot be gained through analysis of an X-ray spectra from a single viewing angle is the hard X-ray directivity, a link to the underlying angular distribution of the accelerated electrons. Through measuring the hard X-ray directivity, the electron distribution can be diagnosed to be isotropic or beamed providing constraints for the acceleration process. Through simultaneously observing a single flare from two different viewpoints at large angles, the X-ray directivity can be determined by comparing the X-ray flux at both viewpoints. For this, hard X-ray spectral analysis is required across both the thermal and non-thermal energy ranges, and hence can be achieved to date with coordinated observations with STIX and the astrophysical mission Fermi/GBM \citep{fermi}. There is also a future opportunity to perform this analysis with cross-calibrated detectors with the proposed PADRE cube-sat mission that will fly STIX spare Caliste-SO detectors to perform this measurement \citep{padre}. These observations will provide the first reliable measurements of and X-ray directivity in flares.

Furthermore, with upcoming hard X-ray instruments with imaging capabilities, such as the Hard X-ray Imager (HXI) on-board the Advanced Space-based Solar Observatory (ASO-S) mission \citep{gan, zhang}, coordinated observations of partially occulted flares from two different viewing angles will allow for both the bright chromospheric hard X-ray sources and the much fainter coronal hard X-ray source to be imaged at the same time \citep{krucker_aso}. As chromospheric hard X-ray footpoints are typically two orders of magnitude brighter than the coronal source, the coronal source is typically only observed during occulted flares when the footpoints are hidden from the field of view behind the solar limb. By combining observations when one instrument sees the flare occulted and the other seeing the flare on-disk, these observations will provide for the first time a full view of all the X-ray emission from a flare, and determine the evolution of the coronal source with the footpoints.

\subsubsection{STIX data access}
The STIX data is available through ESA's Solar Orbiter Archive (SOAR) \footnote{\url{https://soar.esac.esa.int/soar/}} and through the STIX Data Centre \footnote{\url{https://datacenter.stix.i4ds.net}}. More information about the STIX instrument and data products can be accessed at the STIX Data Centre, where also the quick-look data and science data products can be browsed. Software to analyse STIX data is available in IDL through SolarSoftWare (SSW), with some tools also available in Python.

\acknowledgement{ \textit{Solar Orbiter} is a space mission of international collaboration between ESA and NASA, operated by ESA. We thank all the individuals who contributed to STIX, and all the funding agencies that supported STIX: Swiss Space Office, the lead funding agency for STIX, the Polish National Science Centre (grants 2011/01/M/ST9/06096 and 2015/19/B/ST9/02826), Centre national d’études spatiales (CNES), Commissariat à l’énergie atomique et aux énergies alternatives (CEA), the Czech Ministry of Education (via the PRODEX program), Deutsches Zentrum für Luftund Raumfahrt (DLR) (grants: 50 OT 0903, 1004, 1204), the Austrian Space Programme, ESA PRODEX, administered in Ireland by Enterprise Ireland, the Agenzia Spaziale Italiana (ASI) and the Istituto Nazionale di Astrofisica (INAF). L.A.H and S.M. are supported by ESA Research Fellowships. The authors would like to thank Alexander Warmuth and Olivier Limousin for their comments that helped improve this chapter.}

\end{document}